\documentclass[aps,prl,twocolumn,reprint,showpacs,amsmath,amssymb,superscriptaddress,citeautoscript]{revtex4-1}

\usepackage{graphicx}

\begin{document}


\title{Hydrostatic and chemical pressure tuning of CeFeAs$_{1-x}$P$_{x}$O single crystals}

\author{K. Mydeen}
 \email{kamal@cpfs.mpg.de}
 \affiliation{%
Max Planck Institute for Chemical Physics of Solids, N\"{o}thnitzer Str.\ 40, 01187 Dresden, Germany
}%
\author{E. Lengyel}
 \affiliation{%
Max Planck Institute for Chemical Physics of Solids, N\"{o}thnitzer Str.\ 40, 01187 Dresden, Germany
}%
\author{A. Jesche}
 \affiliation{%
Max Planck Institute for Chemical Physics of Solids, N\"{o}thnitzer Str.\ 40, 01187 Dresden, Germany
}%
 \affiliation{%
The Ames Laboratory, Iowa State University, Ames, USA
}%
\author{C. Geibel}
 \affiliation{%
Max Planck Institute for Chemical Physics of Solids, N\"{o}thnitzer Str.\ 40, 01187 Dresden, Germany
}%
\author{M. Nicklas}
 \email{nicklas@cpfs.mpg.de}
  \affiliation{%
Max Planck Institute for Chemical Physics of Solids, N\"{o}thnitzer Str.\ 40, 01187 Dresden, Germany
}%

\date{\today}

\begin{abstract}

We carried out a combined P-substitution and hydrostatic pressure study on CeFeAs$_{1-x}$P$_{x}$O
single crystals in order to investigate the peculiar relationship of the local moment magnetism of
Ce, the ordering of itinerant Fe moments, and their connection with the occurrence of
superconductivity. Our results evidence a close relationship between the weakening of Fe magnetism
and the change from antiferromagnetic to ferromagnetic ordering of Ce moments at $p^*=1.95$\,GPa in
CeFeAs$_{0.78}$P$_{0.22}$O. The absence of superconductivity in CeFeAs$_{0.78}$P$_{0.22}$O and the
presence of a narrow and strongly pressure sensitive superconducting phase in
CeFeAs$_{0.70}$P$_{0.30}$O and CeFeAs$_{0.65}$P$_{0.35}$O indicate the detrimental effect of the Ce
magnetism on superconductivity in P-substituted CeFeAsO.

\end{abstract}

\maketitle

The discovery of superconductivity in LaFeAsO$_{1-x}$F$_{x}$ \cite{Kamihara08} and the highest
$T_c$'s up to 55\,K observed in F-doped SmFeAsO \cite{Chen08,Ren08} have sparked tremendous interest
among the scientific community. In most of the iron-pnictide materials, the application of
hydrostatic pressure or chemical substitution i) introduces superconductivity by suppressing the Fe
spin-density wave (SDW) ordering in the non-superconducting parent compound
\cite{Okada08,Sefat08,Wang09,Zhigadlo11}, ii) induces systematic changes in $T_c$ in pnictide
superconductors \cite{Takahashi08,Zocco08,Hamlin08}.

The Fe moments in CeFeAsO order in a commensurate SDW at about $145$\,K, while the local Ce moments
order antiferromagnetically below $3.7$\,K \cite{Jesche09}. In La(Sm)FeAsO replacement of As by P
results in chemical pressure induced superconductivity \cite{Li10,Wang08}. Here,
CeFeAs$_{1-x}$P$_{x}$O is outstanding among the 1111-type iron-pnictide materials based on rare-earth
elements: P-substitution does also suppress the Fe-SDW ordering, but the rare-earth magnetism of the
Ce moments changes from antiferromagnetic (AFM) at low ($x<0.3$) to ferromagnetic (FM) at higher
P-concentrations ($x\geq0.3$) \cite{Luo10,Jesche12}. Only recently, zero resistance was observed in
CeFeAs$_{0.70}$P$_{0.30}$O single crystals close to the crossover from AFM to FM ordering of the Ce
moments in an already ferromagnetically ordered sample \cite{Jesche12}. The application of pressure
leads to an increase of $T_c$ in doped LaFeAsO$_{1-x}$F$_{x}$ \cite{Takahashi08,Zocco08} and LaFePO
\cite{Hamlin08} pnictide superconductors. On the other hand pressure induces a decrease of $T_c$ in
F-doped \mbox{CeFeAsO} and an enhancement of $T_c$ in Co-doped CeFeAsO \cite{Sun10,Kumar11}. So far
no superconductivity was found in undoped CeFeAsO under pressures up to 50\,GPa \cite{Zocco11}.

In this Letter hydrostatic pressure is used to finetune the physical properties of
CeFeAs$_{1-x}$P$_{x}$O in the pressure region where the type of the Ce ordering changes and
superconductivity has been reported. Application of modest pressures ($p\lesssim3$\,GPa) allows us to
tune CeFeAs$_{0.78}$P$_{0.22}$O through the interesting region. Our investigations evidence a sudden
change of the Ce ordering from AFM to FM at about $p^*\approx1.95$\,GPa, where also the $T_N^{\rm
Fe}(p)$ becomes constant upon further increasing pressures. However, we found no indication for
superconductivity. In CeFeAs$_{0.70}$P$_{0.30}$O and CeFeAs$_{0.65}$P$_{0.35}$O we observed weak
Fe-SDW ordering, FM ordering of Ce moments, and superconductivity at low pressures. External pressure
clearly separates the FM and superconducting (SC) transition temperatures. In the discussion we will
contrast the effect of hydrostatic and chemical pressure.


The details on the preparation and characterization of the single crystalline CeFeAs$_{1-x}$P$_{x}$O
samples can be found in Ref.\,\onlinecite{Jesche10}. In the following we use the nominal P
concentration $x$, which was found to be in good agreement with the actual composition
\cite{Jesche12}. 4-probe electrical-resistance measurements in the $ab$-plane were carried out using
an LR700 resistance bridge. Temperatures down to 1.8\,K and magnetic fields up to 14\,T were achieved
in a Quantum Design PPMS. Magnetic field was applied in the $ab$-plane parallel to the electrical
current. Pressures up to 2.85\,GPa were generated in a double-layer piston-cylinder type pressure
cell using silicon oil as pressure transmitting medium. The pressure shift of the SC transition of
lead served as pressure gauge. The narrow transition at all pressures confirmed the good hydrostatic
conditions inside the pressure cell.

\begin{figure}[tb!]
\includegraphics[angle=0,width=8cm,clip]{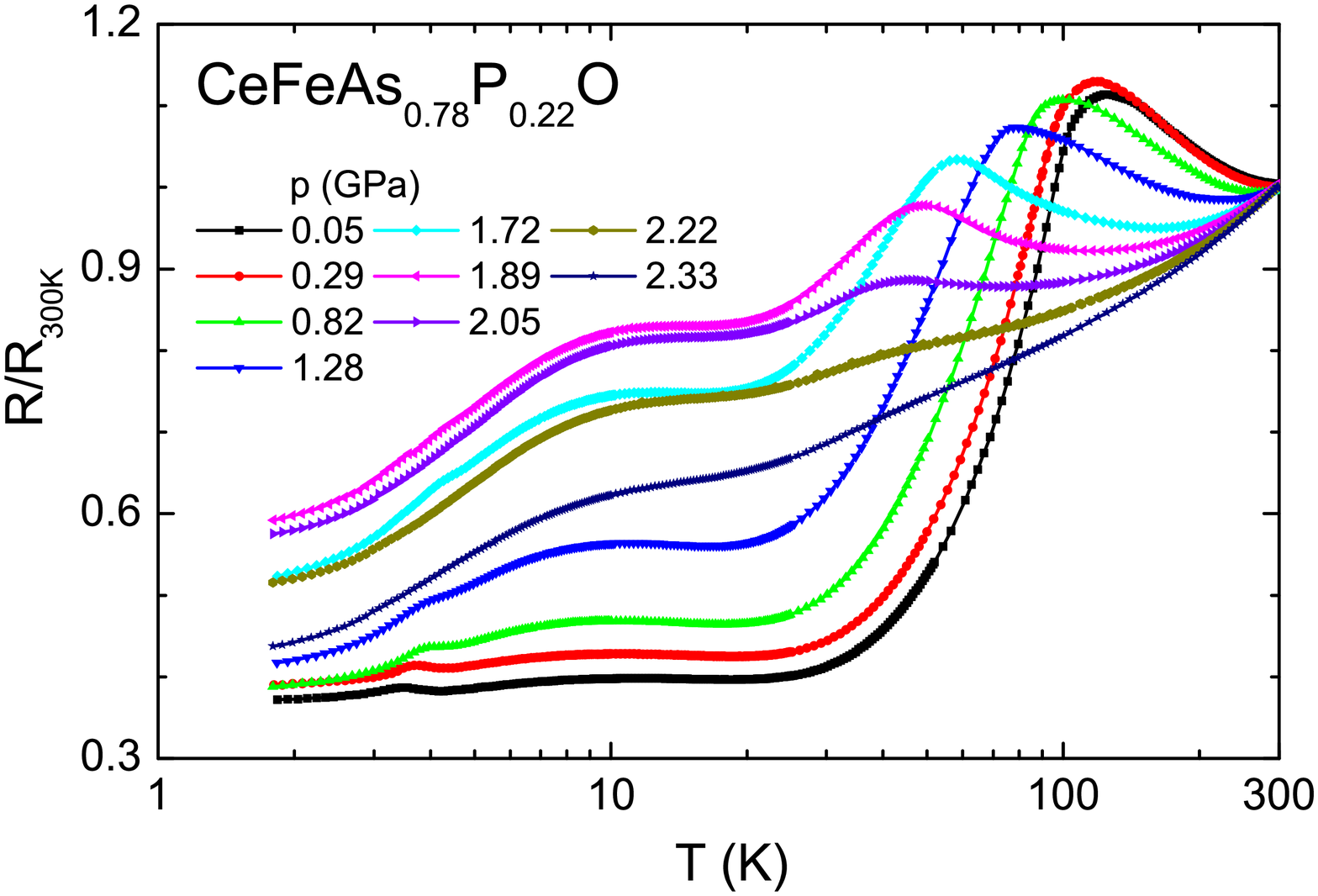}
\caption{\label{Resistance22} (Color online) Temperature
dependence of the electrical resistance of single crystalline
CeFeAs$_{0.78}$P$_{0.22}$O normalized by its value at room
temperature under various hydrostatic pressures up to 2.33\,GPa.}
\end{figure}

The normalized electrical resistance $R(T)/R_\mathrm{300\,K}$ of CeFeAs$_{0.78}$P$_{0.22}$O for
various pressures up to 2.33 GPa is depicted in Fig.\,\ref{Resistance22}. At ambient pressure, $R(T)$
exhibits a maximum followed by a pronounced drop attributed to the onset of Fe-SDW ordering at
$T_N^{\rm Fe} = 93$\,K and a further kink at $T_N^{\rm Ce} = 3.5$\,K due to the AFM ordering of Ce
moments, which is in agreement with Ref.\onlinecite{Jesche12}. Compared with CeFeAsO 22\%
P-substitution already suppresses $T_N^{\rm Fe}$ by about 50\,K. The feature at $T_N^{\rm Fe}$ shifts
to lower temperatures on application of hydrostatic pressure up to $p = 1.95$\,GPa. On further
increasing pressure $T_N^{\rm Fe}(p)$ stays almost constant at about 28\,K ($1.95 {\rm\,GPa} < p <
2.33 {\rm\,GPa}$). It is important to note that the maximum in $R(T)$ becomes sharper and more
pronounced upon increasing pressure for $p\leq1.95$\,GPa, whereas above 1.95\,GPa it broadens and
starts to fade away.

\begin{figure}[tb!]
\includegraphics[angle=0,width=8.5cm,clip]{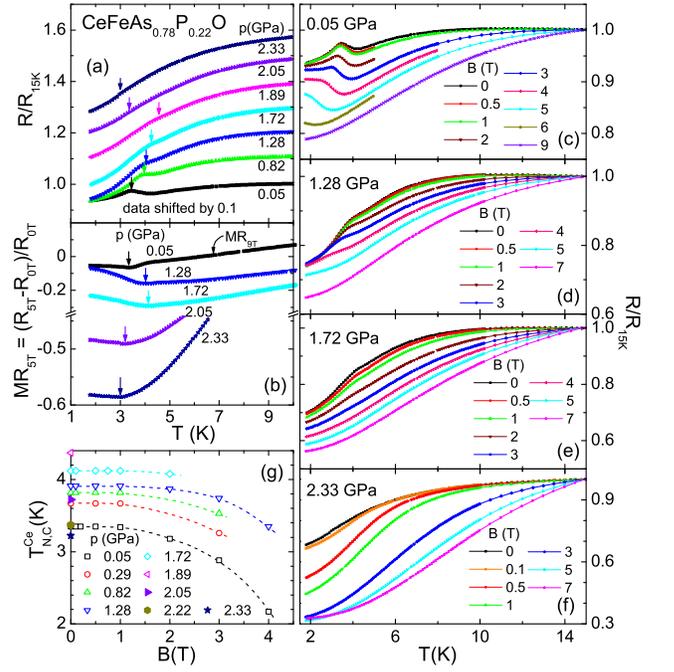}
\caption{\label{lowTresistance} (Color online) a) Temperature dependence of $R(T)$ normalized by its
value at 15\,K of CeFeAs$_{0.78}$P$_{0.22}$O for various pressures up to 2.33\,GPa (shifted by 0.1).
b) Magnetoresistance with magnetic field applied in the $ab$-plane parallel to the electrical current
at different pressures. The arrows in a) and b) mark $T_{N}^{\rm Ce}$ and $T_{C}^{\rm Ce}$, see text
for details. c)-f) Temperature dependence of $R(T)/R_\mathrm{15\,K}$ in different magnetic fields for
$p = 0.05$\,GPa, 1.28 GPa, 1.72 GPa, and 2.33 GPa. Note the different scales for $R/R_\mathrm{15\,K}$
in c)-f). g) Magnetic field dependence of $T_{N,C}^{\rm Ce}$ for various pressures up to 2.33\,GPa.}
\end{figure}

In the following, we focus on the pressure dependence of the Ce ordering. $R(T)/R_\mathrm{15K}$ for
different pressures is shown in Fig.\,\ref{lowTresistance}a. Up to 1.89\,GPa application of pressure
shifts the kink indicating $T_N^{\rm Ce}$ in $R(T)$ to higher temperatures. Upon further increasing
pressure the feature in $R(T)$ broadens significantly and shifts to lower temperatures in contrast to
the behavior at low pressures. While at high pressures the kink at the transition temperature is
hardly visible (we used the maximum in the first derivative of $R(T)$ to determine the transition
temperature), the magnetoresistance MR$_{\rm5T}(T)$ displays a well defined minimum followed by an
increase toward low temperatures due to the Ce ordering (see Fig.\,\ref{lowTresistance}b). The
position of the minimum in MR$_{\rm5T}(T)$ is in good agreement with the results from $R(T)$.
Initially, $T_N^{\rm Ce}(p)$ increases with a rate of about 0.5\,K/GPa, which is only about half the
value reported for polycrystalline CeFeAsO \cite{Zocco11}. We take the abrupt change of the pressure
dependence of the ordering temperature and the significantly broadened feature in $R(T)$ above
1.95\,GPa as a hint for a change of the Ce ordering at $p^*\approx1.95$\,GPa. Comparing our pressure
data with the results of P-substitution in CeFeAs$_{1-x}$P$_{x}$O lets us to propose that the Ce
ordering changes from AFM to FM, which we will substantiate in the following. However, in
CeFeAs$_{0.78}$P$_{0.22}$O under pressure we observe a decrease in FM Ce ordering temperature
($T_C^{\rm Ce}$) above 1.95\,GPa in contrast to the increase of $T_C^{\rm Ce}$ observed on chemical
pressure by phosphorus substitution. This points at differences between the effect of hydrostatic and
chemical pressure on the physical behavior in CeFeAs$_{1-x}$P$_{x}$O, which we will address later.

Electrical resistance measurements in applied magnetic fields give further information on the
magnetic ordering of the Ce moments. Figs.\,\ref{lowTresistance}c-f depict $R(T)/R_\mathrm{15\,K}$ of
CeFeAs$_{0.78}$P$_{0.22}$O in different magnetic fields applied parallel to the $ab$-plane for
selected pressures. At $p = 0.05$\,GPa, a peak-like anomaly in the resistance indicates the AFM
ordering of the Ce moments. Upon increasing magnetic field a more step-like feature develops and
shifts to lower temperatures. The monotonic decrease of $T_N^{\rm Ce}(B)$ is in agreement with the
AFM ordering of the Ce moments in the $ab$-plane. At 1.28\,GPa only a kink in $R(T)$ remains and
marks $T_N^{\rm Ce}$. However, we still observe the same field dependence of $T_N^{\rm Ce}$. Also at
1.72\,GPa we observe a small kink in $R(T)$ shifting to lower temperatures upon increasing $B$. The
effect of a small magnetic field $B\leq0.5$\,T on $R(T)$ is tiny until $p^*\approx1.95$\,GPa. In
contrast, above $p^*$ a huge effect appears above $T_C^{\rm Ce}$ already at low fields. It is worth
mentioning that the value of the change of the resistance in 0.5\,T at 2\,K is about one order of
magnitude larger above $p^*$ than below, substantiating that the Ce magnetism below and above $p^*$
is fundamentally different.

A monotonous decrease of $T_N^{\rm Ce}$ is observed in magnetic
fields for $p \leq p^*$ as shown in Fig.\,\ref{lowTresistance}g.
The $T_N^{\rm Ce}(B)$ curves for different pressures do not
cross; for any fixed magnetic field, $T_N^{\rm
Ce}(p,B)\mid_{B={\rm const}}$ increases upon increasing $p$. Due
to the broadening of the transition anomaly the maximum field up
to which we can define $T_N^{\rm Ce}$ decreases with pressure.
Fig.\,\ref{lowTresistance}f suggests an increase of the ordering
temperature with increasing field, but the curves for $B>0$ do
not present a clear kink allowing for a reliable definition of the
transition temperature. This gives a further hint at the change
of the Ce ordering from AFM to FM (for $p= 1.89$\,GPa no field
data is available).

\begin{figure}[tb!]
\includegraphics[angle=0,width=8cm,clip]{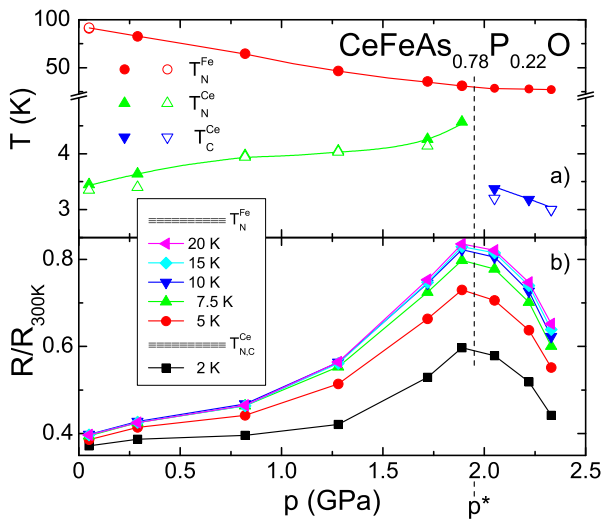}
\caption{\label{Phasediagram22} (Color online) a) $T-p$ phase diagram of CeFeAs$_{0.78}$P$_{0.22}$O.
Solid and open symbols represent the transition temperatures deduced from $R(T)$ and
$\mathrm{MR}(T)$, respectively. b) Pressure dependence of the isothermal resistance normalized by its
value at 300\,K.}
\end{figure}

The $T-p$ phase diagram in Fig.\,\ref{Phasediagram22}a summarizes the results on
CeFeAs$_{0.78}$P$_{0.22}$O. The transition temperatures deduced from electrical resistance and MR are
in good agreement (solid and open symbols, respectively). Upon application of pressure $T_N^{\rm Fe}$
decreases monotonously up to $p^*\approx1.95$\,GPa, upon further increasing pressure $T_N^{\rm Fe}$
is almost constant before its signature in the resistance starts to disappear. $T_N^{\rm Ce}(p)$
monotonously increases with increasing pressure up to $p^* = 1.95$\,GPa. Above $p^*$ our results
indicate a sudden change of the type of the Ce ordering from AFM to FM but, in contrast to
P-substitution in CeFeAs$_{1-x}$P$_{x}$O, $T_C^{\rm Ce}$ decreases upon further increasing pressure.
Up to 2.33\,GPa no indication for superconductivity was found. Even though superconductivity might
develop at higher pressure, we note that we did not find any indication of superconductivity around
$p^*$ where the Ce ordering changes from AFM to FM. In this region superconductivity develops in
chemically pressurized CeFeAs$_{1-x}$P$_{x}$O \cite{Jesche12}.

\begin{figure}[tb!]
\includegraphics[angle=0,width=8cm,clip]{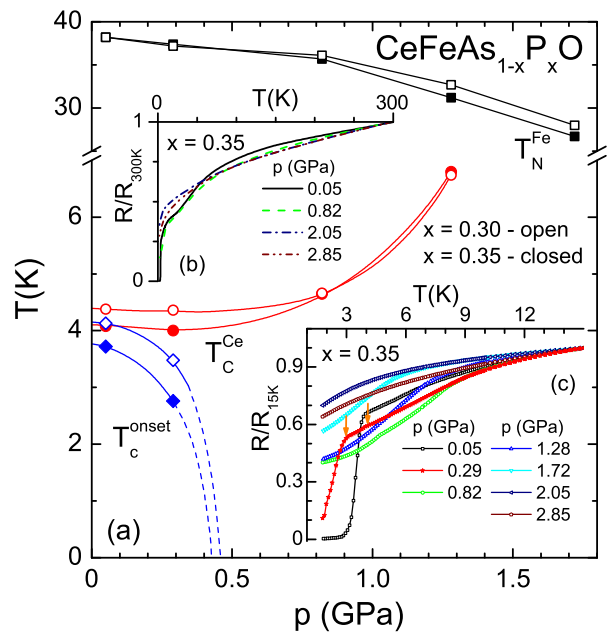}
\caption{\label{Phasediagram3035} (Color online) a) $T - p$ phase diagram for
CeFeAs$_{0.70}$P$_{0.30}$O and CeFeAs$_{0.65}$P$_{0.35}$O. b) $R(T)/R_\mathrm{300\,K}$ for selected
pressures. c) Temperature dependence of the electrical resistance normalized by its value at 15\,K
for various pressures up to 2.85\,GPa. At 0.29~GPa arrows indicate the anomaly at $T_C^{\rm Ce}$ and
the onset of the SC transition at $T_c^\mathrm{onset}$.}
\end{figure}

Fig.\,\ref{Phasediagram22}b displays the normalized isothermal resistance $R_T(p) = R_{T{\rm=
const}}(p)/R_{\rm 300K}(p)$ at different temperatures. The isothermal resistance at 20\,K just below
$T_N^{\rm Fe}$ can be considered as a measure of the strength of the Fe moment fluctuations. $R_{\rm
20\,K}(p)$ possesses a pronounced maximum around 1.95\,GPa, where $T_N^{\rm Fe}(p)$ starts to
saturate. Thus upon increasing pressure the Fe moment fluctuations gradually increase, becoming
strongest around 1.95\,GPa and decrease again for higher pressures. On lowering temperature the
maximum stays at the same pressure. Even more surprising the size of the maximum in $R_T(p)$ remains
almost unchanged upon reducing temperature from 20\,K to 7.5\,K, the latter temperature being far
below $T_N^{\rm Fe}$. This indicates that even at 7.5\,K strong fluctuations of the Fe moments are
present. Furthermore, at 5\,K and even at 2\,K, well below the Ce ordering, a clear maximum shows up
in $R_{T}(p)$. The result at 2\,K is remarkable since it evidences the presence of iron moment
fluctuations down to lowest temperatures. We note that at the same pressure where we observe the
maximum in $R_T(p)$ the Ce ordering changes from AFM to FM.

We now turn to the samples with higher P content.
CeFeAs$_{0.70}$P$_{0.30}$O and CeFeAs$_{0.65}$P$_{0.35}$O show FM
ordering of the Ce moments ($T_C^{\rm Ce}= 4.1$\,K and 4.3\,K,
respectively) and superconductivity at slightly lower
temperatures ($T_c=3.7$\,K and 4.1\,K, respectively). In our
resistance data we detect only a weak anomaly at the SDW
transition at about $T_N^{\rm Fe}$ = 38\,K for both concentrations
in good agreement with Ref.\,\cite{Jesche12}. The results of the
pressure experiments for CeFeAs$_{1-x}$P$_{x}$O, $x = 0.30$ and
0.35, are similar and summarized in the $T - p$ phase diagram in
Fig.\,\ref{Phasediagram3035}a. For both compounds $T_N^{\rm
Fe}(p)$ decreases monotonously upon increasing pressure from
38\,K at ambient pressure to below 30\,K at 1.72\,GPa with an
initial rate ${\rm d}\ln{T_N^{\rm Fe}}/{\rm d}p =
-0.084$\,GPa$^{-1}$. Above this pressure we cannot determine
$T_N^{\rm Fe}$ from the data anymore.

The observed pressure dependence of $T_N^{\rm Fe}(p)$ is surprising considering the previous results
on CeFeAs$_{0.78}$P$_{0.22}$O. In the comparable pressure regime where Ce orders ferromagnetically in
CeFeAs$_{0.78}$P$_{0.22}$O ($p>1.95$\,GPa) $T_N^{\rm Fe}(p)$ is almost constant and much smaller than
in CeFeAs$_{0.70}$P$_{0.30}$O and CeFeAs$_{0.65}$P$_{0.35}$O at ambient pressure. This substantiates
the different effects of chemical and hydrostatic pressure. At low temperatures increasing pressure
effectively separates $T_c(p)$ and $T_C^{\rm Ce}(p)$. In CeFeAs$_{0.65}$P$_{0.35}$O we find that upon
increasing pressure to $p = 0.29$\,GPa, $T_c(p)$ decreases from 3.7\,K at ambient pressure to 2.7\,K,
while the feature at $T_C^{\rm Ce} = 4$\,K becomes clearly visible and separated from $T_c$. Further
increasing pressure above 0.29\,GPa does not leave any signature of superconductivity anymore.
However, we were able to detect $T_C^{\rm Ce}$ up to 1.28\,GPa before we loose the anomaly in our
data. While superconductivity is present, increasing pressure leads to a minute decrease of the
ferromagnetic $T_C^{\rm Ce}$, but once superconductivity disappeared, $T_C^{\rm Ce}$ starts to
increase significantly with increasing pressure in contrast to the smaller P concentration, $x =
0.22$. For both concentrations, $x = 0.30$ and 0.35, $T_c(p)$ is suppressed to zero temperature at an
extrapolated pressure around 0.46\,GPa. It is noticeable that the FM Ce ordering cannot be traced in
electrical resistance data taken in magnetic fields ($B\geq0.1$\,T) parallel to the $ab$ plane (not
shown). This is similar to our findings in CeFeAs$_{0.78}$P$_{0.22}$O at pressures above $p^*$. As
expected, the SC transition shifts toward lower temperatures with increasing the magnetic field. For
$x = 0.35$, the upper critical field $\mu_0H_{c2}^{ab}(0)$ can be estimated to about 1.25\,T for
0.05\,GPa and only 0.25\,T for 0.29\,GPa taking $T_c^\mathrm{onset}(H)$, indicating that the value of
$H_{c2}^{ab}(0)$ is more effectively suppressed by pressure than $T_c(p)$.


In the layered iron pnictides, hydrostatic and chemical pressure reduces the ratio of the $c$- and
$a$-axis lattice parameters, which is resulting in a strong influence on the antiferromagnetically
ordered Fe moment and $T_N^{\rm Fe}$ \cite{Cruz10,Sefat11}. The chemical pressure by P-substitution
in \textit{Ln}FeAsO (\textit{Ln}=Ce, La, Sm) compresses the $c$-axis stronger than the $a$-axis, and
furthermore decreases the pnictogen (\textit{Pn}) height
\cite{Cruz10,Wang09,Luo10,Kumar11,Zhigadlo11}. This corresponds to a strong compressing of the FeAs/P
layer and therefore one expects a strong increase of the hybridization between Fe and \textit{Pn}
states. Accordingly we observe a strong initial decrease of $T_N^{\rm Fe}$ with hydrostatic pressure
by ${\rm d}\ln{T_N^{\rm Fe}}/{\rm d}p = -0.36$\,GPa$^{-1}$ in CeFeAs$_{0.78}$P$_{0.22}$O compared
with only $-0.071$\,GPa$^{-1}$ in CeFeAsO \cite{Zocco11}. In contrast, chemical pressure results in a
stretching of the \textit{Ln}O layer \cite{Wang09}, while hydrostatic pressure results in a
compressing of this layer \cite{Kumar11}. This is very likely the reason for the different dependence
of $T_N^{\rm Ce}$ and $T_C^{\rm Ce}$ under hydrostatic and chemical pressure.

In summary, we carried out a hydrostatic and chemical pressure investigation on CeFeAs$_{1-x}$P$_x$O
single crystals. In CeFeAs$_{0.78}$P$_{0.22}$O we found first a fast decrease of $T_N^{\rm Fe}(p)$
and then above $p^*=1.95$\,GPa a leveling of at about 28\,K. This behavior seems to exclude a quantum
critical point scenario in CeFeAs$_{0.78}$P$_{0.22}$O under pressure. At $p^*$ the Ce ordering
changes from AFM to FM. Our analysis of the isothermal resistivity suggests the presence of Fe moment
fluctuations down to lowest temperatures. We notice that the magnetic ordering of the Ce changes from
AFM to FM ordering at the same pressure where we find the maximum in the isothermal resistance. We do
not find superconductivity in the region around $p^*$ in contrast to the results of P-substitution in
CeFeAs$_{1-x}$P$_x$O where superconductivity is observed coexisting with ferromagnetism close to the
P-concentration where the AFM Ce ordering changes to FM \cite{Jesche12}. The reason could lie in the
different response of CeFeAs$_{0.78}$P$_{0.22}$O to pressure and to P-substitution as it is
evidenced, for example, in the different behavior of $T_{N,C}^{\rm Ce}$ on pressure and
P-substitution. However, we point out that we cannot exclude the appearance of superconductivity at
pressure higher than our experimentally accessible range. CeFeAs$_{0.70}$P$_{0.30}$O and
CeFeAs$_{0.65}$P$_{0.35}$O are situated right in the narrow P-concentration regime where
superconductivity and FM ordering of Ce moments coexist. At ambient pressure we observe a weak Fe-SDW
ordering around $T_N^{Fe}=38$\,K which is suppressed upon increasing pressure. In both compounds
increasing pressure enhances $T_C^{\rm Ce}(p)$. The superconductivity is highly sensitive to pressure
and $T_c(p)$ is already suppressed to zero temperature around $p\approx0.46$\,GPa. Our study
highlights the delicate interplay between iron and cerium magnetism, their sensitivity to structural
properties and, not at last, the subtle connection to superconductivity.

This work was supported by the DFG within the framework of the SPP1458.

\end{document}